\documentclass[a4paper]{aa}
\usepackage{times}
\usepackage{amssymb}
\usepackage{graphicx}
\begin{document}
\title{Stretching of the toroidal field and generation of magnetosonic
       waves in differentially rotating plasma} 
\author{A.~Bonanno\inst{1,2}, V.~Urpin\inst{1,3}}
\offprints{}
\institute{$^{1)}$ INAF, Osservatorio Astrofisico di Catania,
           Via S.Sofia 78, 95123 Catania, Italy \\
           $^{2)}$ INFN, Sezione di Catania, Via S.Sofia 72,
           95123 Catania, Italy \\
           $^{3)}$ A.F.Ioffe Institute of Physics and Technology and
           Isaac Newton Institute of Chile, Branch in St. Petersburg,
           194021 St. Petersburg, Russia}

\date{\today}

\abstract
{}
{We evaluate the generation of magnetosonic waves in differentially rotating
magnetized plasma.} 
{Differential rotation leads to an increase of the azimuthal field by winding 
up the poloidal field lines into the toroidal field lines. An amplification of weak 
seed perturbations is considered in this time-dependent background state.} 
{It is shown that seed perturbations can be amplified by several orders of
magnitude in a differentially rotating flow. The only necessary condition for 
this amplification is the presence of a non-vanishing component of the 
magnetic field in the direction of the angular velocity gradient.}
{}  

\authorrunning{A.Bonanno, V. Urpin}
\titlerunning{Generation of magnetosonic waves in rotating plasma}

\maketitle

\section{Introduction}
In astrophysical bodies, differential rotation is often associated with  
magnetic fields of various strength and geometry. If the poloidal field has 
a component parallel to the gradient of angular velocity, then differential 
rotation can stretch toroidal field lines from the poloidal ones. In the 
presence of the magnetic field, differential rotation can be the reason for 
various magnetohydrodynamic instabilities, particularly if the field geometry
is complex. Some of these instabilities occur in the incompressible limit 
(Velikhov 1959; Fricke 1969; Acheson 1978; Balbus \& Hawley 1991) 
which applies if the magnetic field is weak and the Alfv\'en velocity is 
smaller than the sound speed. Other instabilities become important only in sufficiently 
strong magnetic fields when the effect of compressibility plays a significant 
role (Pessah \& Psaltis 2005; Bonanno \& Urpin 2006, 2007). Note 
that incompressible instabilities can often be suppressed by a strong magnetic
field. For example, the magnetorotational instability (MRI) does not occur if 
the magnetic field satisfies the condition $B^2 > -8 \pi \rho s \Omega \Omega'
L^2$ where $\rho$ is the density, $L$ is the lengthscale of disturbances, 
$\Omega=\Omega(s)$ is the angular velocity, and $s$ is the cylindrical radius;
$\Omega'= d \Omega/ds$ (see, e.g., Urpin 1996, Kitchatinov \& R\"udiger 1997).
Moreover, if $\Omega$ increases with the cylindrical radius MRI cannot arise.
 
In recent years, many simulations of differentially rotating magnetized 
bodies have been performed, and much of the dynamics was interpreted as being a 
direct consequence of the MRI (Brandenburg et al. 1995; Hawley at al. 
1995; Matsumoto \& Tajima 1995; Hawley 2000). Obviously, the 
MRI cannot be the only instability that operates in a rotating magnetized gas. 
For example, stratification can lead to a number of strong non-axisymmetric 
instabilities (Agol et al. 2001; Narayan et al. 2002; Keppens et al. 2002). 
Blokland et al. (2005) consider the influence of a toroidal 
field on the growth rate of the MRI and find that it leads to overstability 
(complex eigenvalue). Van der Swaluw et al. (2005) study the interplay between 
different instabilities and argue that the growth rate of convection can be 
essentially increased due to magnetorotational effects. Note, however, that 
these studies treat the stability of the magnetic field with a vanishing 
radial component, a condition which is often not met in astrophysical bodies. 
In fact, the presence of a radial magnetic field can change substantially 
the stability properties (Bonanno \& Urpin 2006, 2007). 

In this paper, we consider stability of a differentially rotating gas in the
presence of a non-vanishing radial magnetic field. Differential rotation 
causes the azimuthal field to increase with time by winding up the poloidal 
field lines into the toroidal ones. Therefore, a development of small 
perturbations occurs in the time-dependent background state. We show that 
stretching of the azimuthal field leads to the generation of magnetosonic
waves in a flow. Magnetohydrodynamic waves and turbulence generated by this
instability can play an important role in enhancing transport processes in 
various astrophysical bodies, such as accretion and protoplanetary Disks, 
galaxies, stellar radiative zones, etc. 

\section{Basic equations}

The equations of compressible MHD read  
\begin{eqnarray}
\dot{\vec{v}} + (\vec{v} \cdot \nabla) \vec{v} = - \frac{\nabla p}{\rho} 
+ \vec{g}  + \frac{1}{4 \pi \rho} (\nabla \times \vec{B}) \times \vec{B}, 
\end{eqnarray}
\begin{equation}
\dot{\rho} + \nabla \cdot (\rho \vec{v}) = 0, 
\end{equation}
\begin{equation}
\dot{\vec{B}} - \nabla \times (\vec{v} \times \vec{B}) = 0,
\end{equation}
\begin{equation}
\nabla \cdot \vec{B} = 0. 
\end{equation} 
Our notation is as follows: $\rho$ and $\vec{v}$ are the gas density and 
velocity, respectively; $p$ is the pressure; $\vec{B}$ is the magnetic field 
and $\vec{g}$ is gravity. {In this paper, we consider an isothermal gas
and assume 
\begin{equation}
p = c^2 \rho,
\end{equation}
where the sound speed,  $c^2$,  is constant.}  

We work in cylindrical coordinates ($s$. $\varphi$, $z$) with the unit 
vectors ($\vec{e}_{s}$, $\vec{e}_{\varphi}$, $\vec{e}_{z}$). The basic state 
on which the stability analysis is performed is assumed to be axisymmetric 
with the angular velocity $\Omega = \Omega(s)$ and $\vec{B} \neq 0$. In the
presence of the non-vanishing radial field $B_s$ and differential rotation, the 
azimuthal field increases with time by winding up the radial field lines. If 
the magnetic Reynolds number is large, then one obtains from Eq.~(4) that the 
azimuthal field grows linearly with time in the basic state,
\begin{equation}
B_{\varphi}(t) = B_{\varphi}(0) + s \Omega' B_{s} t,
\end{equation} 
where $B_{\varphi}(0)$ is the azimuthal field at $t=0$. 
{ A growth of $B_{\varphi}$ given by Eq.~(6) can last only while diffusion
of the toroidal field is negligible. Eventually, in the presence of a finite 
diffusivity, a steady state will emerge where winding up is balanced by 
diffusion of the azimuthal field. The time-scale to reach this steady state 
is approximately equal to the diffusion time-scale, $\sim s^2 / \eta$, where 
$\eta$ is the magnetic diffusivity. Therefore, our consideration
is valid only for $t < s^2 / \eta$. However, for astrophysical plasmas, this
time can be very long because of low $\eta$ and large length-scale.}

For the sake of 
simplicity, we assume that gravity is radial, $\vec{g}(s) = -g(s) \vec{e}_s$, 
and the basic state is in hydrostatic equilibrium, then 
\begin{equation}
\frac{\nabla p}{\rho} = \vec{D} + \frac{1}{4 \pi \rho} 
(\nabla \times \vec{B}) \times \vec{B} \;\;, \;\;\;\;
\vec{D} = \vec{g} + \Omega^{2} \vec{s}.
\end{equation}
We consider the basic state in which $B_{s} \propto B_{\varphi}(0) \propto
B_{\varphi}(t) \propto 1/s$. It follows from Eq.~(6) that this condition is
satisfied if $s \Omega' =$const. Then, $\nabla \times \vec{B}(t) =0$ and the
Lorentz force is vanishing in Eq.~(7). If, additionally, the centrifugal 
force is balanced by gravity, $g(s)=s \Omega^2$, then the pressure and 
hence, density are homogeneous in the basic state.

Consider stability of axisymmetric short wavelength perturbations with the 
spatial dependence $\propto \exp ( - i \vec{k} \cdot \vec{r})$ where 
$\vec{k}$ is the wavevector. Small perturbations will be indicated by 
subscript 1, while unperturbed quantities will have no subscript. For the 
purpose of illustration, we treat in this paper the simplest case when 
$\vec{k}$ has only the vertical component, $\vec{k}=(0, 0, k)$. Then, the 
condition $\nabla \cdot \vec{B}_1 =0$ yields $B_{1z}=0$. {The other 
non-trivial linearized MHD-equations to lowest order in $|\vec{k} \cdot 
\vec{r}|^{-1}$ read}  
\begin{equation}
\frac{d \vec{v}_{1z}}{d t}     
= \frac{i k  p_{1}}{\rho}  
+ \frac{i}{4 \pi \rho} \; k (\vec{B}_{1} \cdot \vec{B}) ,  
\end{equation}
\begin{equation}
\frac{d \rho_{1}}{d t} - i \rho k v_{1z} = 0, 
\end{equation}
\begin{equation}
\frac{d B_{1s}}{dt} = i B_s k v_{1z}, 
\end{equation}
\begin{equation}
\frac{d B_{1 \varphi}}{d t} = s \Omega' B_{1s } + 
i k B_{\varphi}(t) v_{1z}. 
\end{equation}
Since gas is assumed to be isothermal, we have $p_1/p = \rho_1/\rho$. 
Eliminating all variables from Eqs.~(8)-(11) in favour of the vertical 
velocity perturbation $v_{1z}$ one obtains the following fourth-order ordinary 
differential equation in $v_{1z}$:
\begin{eqnarray}
\frac{d^4 v_{1z}}{d t^4} + k^2 [c_s^2 + c_A^2(t)] \frac{d^2 v_{1z}}{d t^2}
+ 6 \omega_{B \Omega}^3 \frac{d v_{1z}}{d t}
\nonumber \\
+ 6 k^2 c_{As}^2 (s \Omega')^2 v_{1z} = 0,
\end{eqnarray} 
where
\begin{eqnarray}
\lefteqn{  c_A^2(t) = c_{As}^2 + c_{A \varphi}^2 (t)
\;, \;\; \vec{c}_{As} = \vec{B}_{s} /\sqrt{4 \pi \rho} \;,} \nonumber \\
\lefteqn{\vec{c}_{A \varphi}(t) = \vec{B}_{\varphi}(t) /\sqrt{4 \pi \rho} \;, 
\;\;
\omega^{3}_{B \Omega} = k^{2} c_{A \varphi}(t) c_{A s} s \Omega'.} 
\nonumber.
\end{eqnarray}
Eq.~(12) with the corresponding initial conditions describes the behaviour
of velocity perturbations on the time-dependent background state. The 
behaviour of perturbations of the radial and azimuthal magnetic field can then be 
calculated  from Eq.~(10) and (11).

\section{Numerical results}
To follow the behaviour of perturbations, it is convenient to 
introduce the dimensionless quantities
\begin{eqnarray}
\lefteqn{\tau = \Omega t \;,\;\; q = \frac{s \Omega'}{\Omega} \;,\;\;
x = k H \;, \;\; H = \frac{c_s}{\Omega} \;, } 
\nonumber \\
\lefteqn{\beta_s = \frac{ c_{As}^2 }{c_s^2} \;,\;\;
\beta_{\varphi} = \frac{c_{A \varphi 0}^2 }{c_s^2} \;,\;\;
c_{A \varphi 0} = \frac{B_{\varphi}(0)}{\sqrt{4 \pi \rho}}.} 
\nonumber 
\end{eqnarray}
Then, introducing $v= v_{1z}(t)/v_{1z}(0)$ where $v_{1z}(0)$ is the initial
velocity perturbation, we obtain from Eq.~(12) 
\begin{eqnarray}
\frac{d^4 v}{d \tau^4} + x^2 [1 + \beta_s + (\sqrt{\beta_{\varphi}}  
+ q \sqrt{\beta_s} \tau)^2] \frac{d^2 v}{d \tau^2} +
\nonumber \\
6 q x^2 \sqrt{\beta_s} (\sqrt{\beta_{\varphi}} + q \sqrt{\beta_s} \tau)
\frac{d v}{d \tau}
+ 6 q^2 x^2 \beta_{s} v = 0.
\end{eqnarray}
The dependence of the solution on the wavelength is characterized by the 
parameter $x$ and, on the initial magnetization of gas, by the parameters 
$\beta_s$ and $\beta_{\varphi}$. To solve Eq.~(13), one needs the initial
conditions for three time derivatives of $v$. In actual calculations we try 
different initial conditions because their choice is determined by the 
origin of perturbations which is uncertain. 

Eqs.~(10)-(11) can also be written in a dimensionless form. We have
\begin{eqnarray}
\frac{d C_{As}}{d \tau} = i x \sqrt{\beta_s} v(\tau), \\
\frac{d C_{A \varphi}}{d \tau} = q C_{As} + i x (\sqrt{\beta_{\varphi}} +
q \sqrt{\beta_s} \tau) v(\tau), 
\end{eqnarray}
where
\begin{equation}
C_{As} = \frac{B_{1s}(\tau)}{\sqrt{4 \pi \rho} v_{1z}(0)} \;, \;\;\;
C_{A \varphi} = \frac{B_{1 \varphi}(\tau)}{\sqrt{4 \pi \rho} v_{1z}(0)}.
\end{equation}
In calculations, we assume that the initial perturbations of the magnetic 
field are vanishing. Eqs.~(13)-(15) were solved numerically for a wide range 
of the parameters. 

\begin{figure}
\begin{center}
\includegraphics[width=9.0cm]{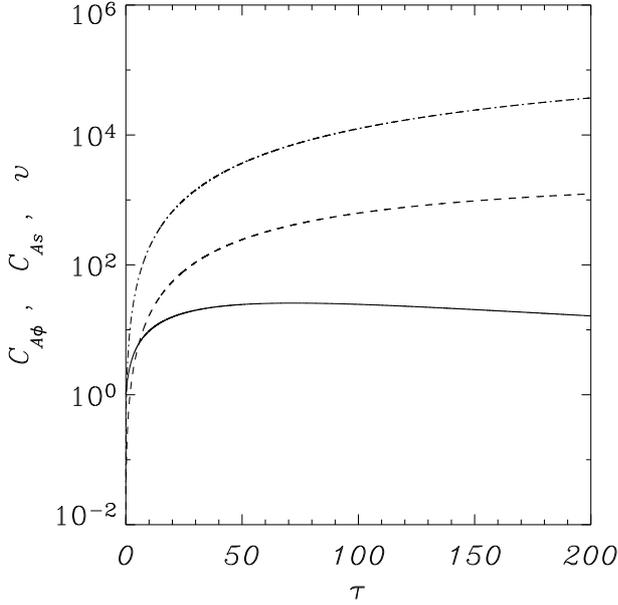}
\caption{The time dependence of $v$ (solid line), $C_{As}$ (dashed 
line), and $C_{A \varphi}$ (dashed-and-dotted line) for $q=0.1$, $x= 3$, 
$\beta_s=0.01$, and $\beta_{\varphi}=1$.}
\end{center}
\end{figure}

In Fig.~1, we plot the time dependence of $v$, $C_{As}$, and $C_{A \varphi}$ 
in the regions where the angular velocity increases slowly with $s$ ($q= 
0.1$). The other parameters are $x=3$, $\beta_s=0.01$, and $\beta_{\varphi}= 
1$. Integrating Eq.~(13), we choose the following initial conditions:
$d v(0)/d \tau = 1$ and $d^2 v(0)/d \tau^2 = d^3 v(0)/d \tau^3 =0$. The 
behaviour of the velocity and magnetic field turns out to be substantially 
different. During the initial stage, the velocity reaches a flat maximum at 
$\tau \sim 60-70$ which corresponds approximately to $10 P$ where $P$ is the 
rotation period. At that time, the perturbation of velocity is $\approx 25$ 
times greater than its initial value. Perturbations of the magnetic field 
are much larger. After $\tau \sim 100$ ($t \sim 15 P$), $C_{As}$  
approaches  the saturation level that is approximately a factor $10^3$ 
greater than the initial perturbation of $v_{1z}$. The amplification of the 
azimuthal field is even larger. After $t \sim 15 P$, $C_{A \varphi}$ is a 
factor $10^4$ stronger than the initial perturbation $v_{1z}(0)$, and it 
continues to grow approximately linearly with time. Note that the long-term 
behaviour ($\tau \gg 200$) of these quantities is also different. The velocity
$v$ changes the sign and exhibits some sort of oscillatory behaviour with the 
amplitude growing $\propto t^{1/2}$ whereas $C_{As}$ reaches the saturation 
level. A perturbation of the azimuthal field $B_{1 \varphi}$ is still 
$\propto t$ at very large $t$.   

\begin{figure}
\begin{center}
\includegraphics[width=9.0cm]{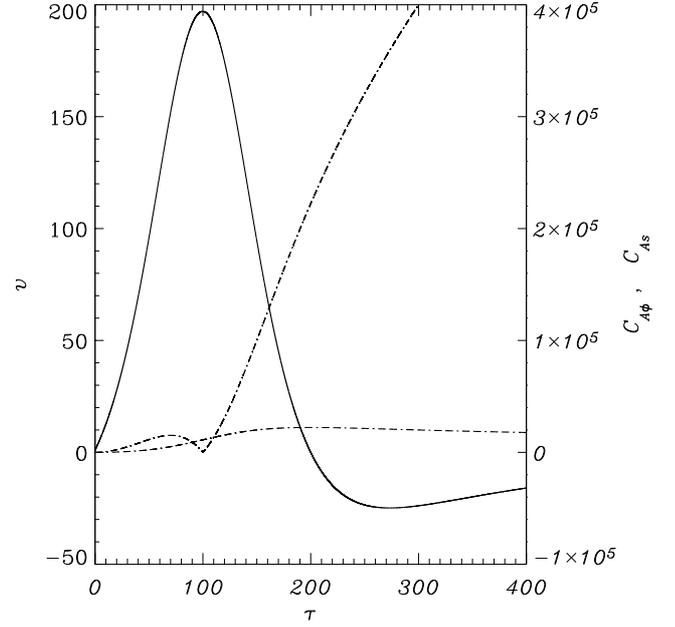}
\caption{Same as in Fig.~1 but for $q= -0.1$ and $x=10$.}
\end{center}
\end{figure}

In contrast with the magnetorotational instability, the behaviour of 
perturbations is not crucially sensitive to the sign of $\Omega'$. To 
illustrate this point, we show in Fig.~2 the time dependence in the case $q= 
- 0.1$ ($\Omega$ decreases slowly with $s$). Qualitatively, the 
dependences are the same. The only essential difference is that the azimuthal 
field generated by differential rotation has a negative sign because the 
radial gradient of the angular velocity is negative. Initially ($\tau <
100$), the main contribution to $B_{1 \varphi}$ is provided by compressibility
(the second term on the r.h.s. of Eq.~(11)).  Later on, however  the azimuthal 
magnetic perturbations are determined by winding up the radial field 
perturbations
and becomes negative. In Fig.~2, the modulus of $C_{A \varphi}$ is shown.
Since $x$ is larger in this figure, perturbations of all quantities can reach 
higher values compared to Fig.~1. For instance, the velocity becomes $\sim 
200$ times greater than the initial value after 15 rotation periods. The radial 
perturbation of the Alfv\'en velocity reaches the saturation level which is 
$\sim 2 \times 10^4$ times greater than $v_{1z}(0)$, and the azimuthal 
perturbation is even about 10 times larger. For very long $t$ ($\tau \gg 
200$), the velocity exhibits the oscillatory behaviour with the 
amplitude $\propto t^{1/2}$. 
\begin{figure}
\begin{center}
\includegraphics[width=9.0cm]{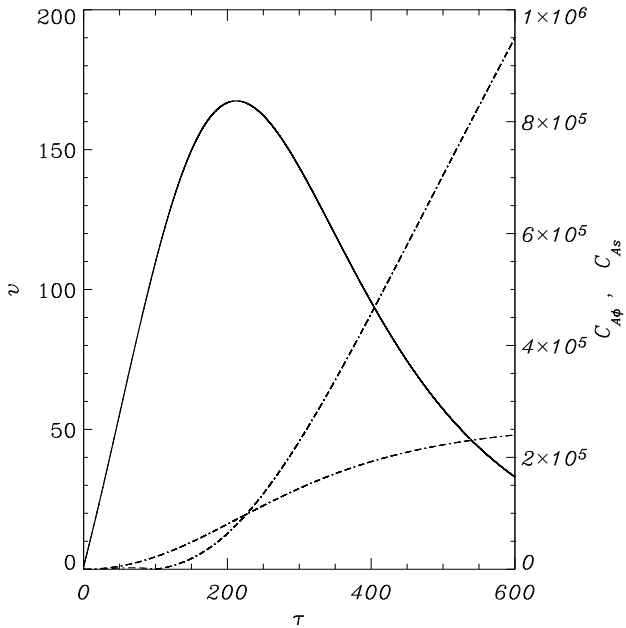}
\caption{Same as in Fig.~2 but for $q=-0.01$, $\beta_s = \beta_{\varphi} =
0.1$, and $x=10$. The initial conditions are as for in Fig.~1.}
\end{center}
\end{figure}
In Fig.~3, we plot the behaviour of perturbations in the case of a very
weak differential rotation, $q = 0.01$. For a weak rotation shear, the 
amplification of initial perturbations is slower. The velocity reaches the
maximum after $\sim 40 P$, and this maximum is rather high: $v_{1z}$ exceeds
its initial value by a factor $\sim 100$. Despite a slower growth rate,
saturation of $C_{As}$ still occurs at a rather high level ($C_{As} \sim
2 \times 10^4$). As expected, the azimuthal field grows linearly with time and 
can reach a very high value. 

\begin{figure}
\begin{center}
\includegraphics[width=9.0cm]{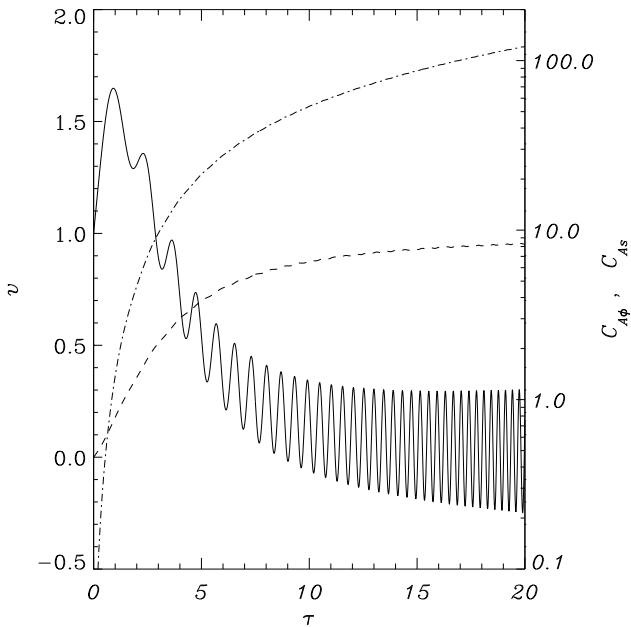}
\caption{Same as in Fig.~1 but for $q=1$ and $\beta_{s}=\beta_{\varphi}=0.1$.}
\end{center}
\end{figure}

The same dependences, for a relatively large value of $q$,  are  shown in 
Fig.~4. In this case, perturbations exhibit a clear oscillatory behaviour. 
This concerns particularly perturbations of the vertical velocity. After a 
short initial relaxation (lasting $\sim 2-3$ rotation period), $v$ reaches 
a self-similar regime when the amplitude of oscillations grows with time  
$\propto \tau^{1/2}$ and the frequency $\propto \tau$. Perturbations of the 
magnetic field also exhibit a weak oscillatory behaviour but the amplitude of 
oscillations is much smaller. This is because the field components are given 
by the integrals of a rapidly oscillating function $v$ (see Eq.~(10) and 
(11)), and oscillations in $B_{1s}$ and $B_{1 \varphi}$ are smoothed. 
Perturbations of the radial field reaches a saturation level after $\sim 2-3 
P$. In saturation, the radial Alfv\'en velocity $C_{As}$ is approximately 10 
times greater than the initial perturbation of the vertical velocity. 
Perturbations of the azimuthal field grow rather rapidly at the initial stage 
however, later on (at $\tau >2 P$), they continue to grow linearly with time.  

\begin{figure}
\begin{center}
\includegraphics[width=9.0cm]{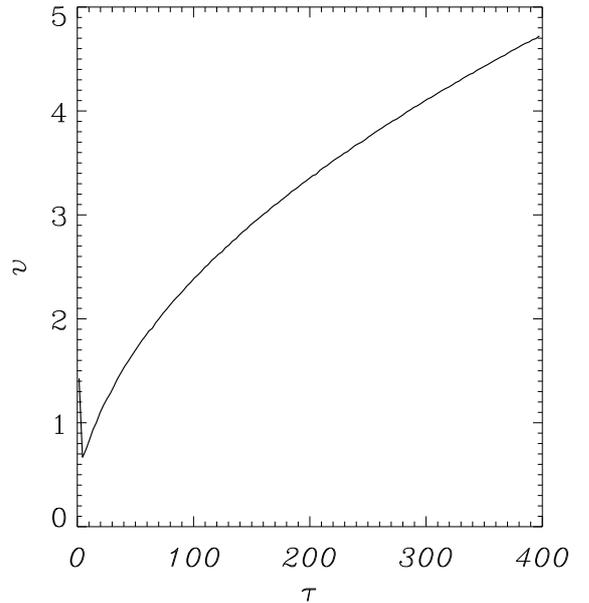}
\caption{The long-term behaviour of the amplitude of $v$ for $\beta_{s}=
\beta_{\varphi}=0.1$, $q=2$, and $x=3$. }
\end{center}
\end{figure}

In Fig.~5, we show the long-term behaviour of the amplitude of the vertical
velocity. Except for the initial stage, the numerical results are well fitted by
the same dependence $\propto \tau^{1/2}$ as that in Fig.~4. We can compare these 
numerical results with the analytical solution of Eq.~(13) for large $\tau$. 
It can be easily checked  that the asymptotic solution of Eq.~(13) at large 
$\tau$ is given by
\begin{equation}
v \propto \sqrt{\tau} \cos \left( \frac{1}{2} q x \sqrt{\beta_s} \tau^2 + 
\frac{\pi}{8} \right) + O(\tau^{-3/2}).
\end{equation}  
Indeed, it can be noted that the vertical velocity is $\propto \tau^{1/2}$ 
and the frequency of oscillations is $\propto \tau$ at $\tau \rightarrow 
\infty$ in complete agreement with our numerical results. The frequency of 
oscillations depends on 
the magnetic field and is higher for a stronger initial radial field. Note 
that the asymptotic solution (17) is valid for a wide range of parameters and 
initial conditions. It is clear from  solution (17) that $B_{1s}$ should reach
saturation quite rapidly. We have from Eq.~(14) that
\begin{equation}
C_{As}(\tau) = i x \sqrt{\beta_{s}} \int^{\tau}_{0} v(\tau') d \tau'. 
\end{equation}
The time integral of the asymptotic part of $v$ (Eq.~(17)) gives a negligible 
contribution to $C_{As}$, and the only non-vanishing contribution can be 
provided by the value of $v$ at relatively short $\tau$. If $C_{As}$ reaches 
saturation rapidly then, as it follows from Eq.~(15), $C_{A \varphi}$ should
grow approximately $\propto \tau$ at large $\tau$. This sort of behaviour
agrees perfectly with the results of our numerical calculations.

\section{Discussion}

We have shown that the winding up of toroidal field lines from the poloidal field lines
is accompanied by an amplification of seed perturbations of the velocity and
magnetic field. A very simplified model has been considered in this paper, 
but we believe that qualitatively the same results can be obtained for more 
general background states and perturbations. Differential rotation and 
compressibility of the gas lead to the generation of magnetosonic waves with 
the amplitude that grows with time. The physical processes responsible for 
this amplification are exactly the same that result in the instability 
considered by Bonanno \& Urpin (2006, 2007). The only difference is that, 
in this paper, we consider the development of perturbations on a 
time-dependent background state and, as a result, the growth of 
perturbations is not exponential. 

The behaviour of seed perturbations depends essentially on various parameters 
and can generally be rather complex. If the parameter $q=s \Omega'/\Omega$ 
is relatively small ($\leq 0.1$), then the perturbations of velocity and 
magnetic field initially grow monotonously and can reach quite high values.
For example, the perturbation of the vertical velocity becomes approximately 
$150-200$ times greater than its initial value after only 15-30 rotation 
periods (see Figs.~2 and 3). Perturbations of the magnetic field reach even 
higher values during the initial stage. For instance, the Alfv\'en velocity 
corresponding to the perturbation of the radial field component, $C_{As}$, 
can exceed the initial velocity perturbation by a factor $\sim (1-3) \times 
10^5$ after the same time, but the perturbations of the toroidal field are 
even stronger. As a result of such a strong initial amplification, seed
perturbations can already reach a non-linear regime  after 15-30 rotation 
periods if their initial values are sufficiently large. Further 
evolution of perturbations will then be entirely determined by non-linear effects.
However, if the non-linear regime is not reached during this initial stage, the
behaviour of perturbations becomes oscillatory with slowly growing amplitude
($\propto t^{1/2}$). At sufficiently large $t$, the frequency of oscillations
grows linearly with time and is given approximately by
\begin{equation}
\omega \approx \Omega \left( \frac{1}{2} k H \sqrt{\beta_s} s \Omega' t
\right).
\end{equation} 
Note that perturbations of the magnetic field exhibit more regular behaviour
because they can be expressed in terms of the time integrals of a rapidly 
oscillating velocity. In the case of a strong differential rotation ($q \sim 
1$), perturbations exhibit the oscillatory behaviour from the very beginning 
and the initial growth of their amplitude is less significant.  

The generation of magnetosonic waves occurs even if the magnetic field is 
very strong and suppresses different MHD-instabilities which can arise in a 
differentially rotating flow (for example, the MRI). The presence of
differential rotation and radial magnetic field is, however, crucially 
important for the considered process. Since both differential rotation and
radial field are quite common in astrophysics, we believe that the
considered mechanism can occur in various astrophysical bodies and plays an 
important role in enhancing transport processes in plasma.

\vspace{0.5cm}
\noindent
{\it Acknowledgments.}
This research project has been supported by a Marie Curie Transfer of
Knowledge Fellowship of the European Community's Sixth Framework
Programme under contract number MTKD-CT-002995. 
VU thanks INAF-Ossevatorio Astrofisico di Catania for hospitality.

{}


\begin{thebibliography}{}

\item{}
Acheson D. 1978. Phil. Trans. R. Soc. London, 289A, 459

\bibitem{}
Agol E., Krolik J., Turner N., Stone J. 2001. ApJ, 558, 543

\item{}
Balbus S.A., \& Hawley J.F. 1991. ApJ, 376, 214 

\bibitem{}
Blokland J.W.S., van der Swaluw E., Keppens R., Goedbloed J.P. 2005.
A\&A, 444, 337

\bibitem{}
Bonanno, A. \& Urpin, V. 2006. Phys. Rev. E, 73, 066301

\bibitem{}
Bonanno, A. \& Urpin, V. 2007. ApJ, 662, 851

\bibitem{}
Brandenburg, A., Nordlund, \AA., Stein, R., Torkelsson, U., 1995, ApJ,
446, 741 

\bibitem{}
Hawley, J. 2000. ApJ, 528, 462

\bibitem{}
Hawley, J., Gammie, C., \& Balbus, S. 1995. ApJ, 440, 742

\bibitem{}
Fricke K. 1969. A\&A, 1, 388

\item{}
Keppens, R., Casse, F., \& Goedbloed, J. 2002, ApJ,  569, L121

\item{}
Kitchatinov, L.L., \& R\"{u}diger, G. 1997, MNRAS,  286, 757

\item{}
Matsumoto, R., \& Tajima, T. 1995, ApJ,  445, 767 

\bibitem{}
Narayan R., Quataert E., Igumenshchev I., Abramowicz M. 2002. ApJ, 577, 295

\bibitem{}
Pessah, M. \& Psaltis, D. 2005. ApJ, 628, 879 

\bibitem{}
Urpin, V. 1996. MNRAS,  280, 149

\bibitem{}
Van der Swaluw E., Blokland J.W.S., \& Keppens R. 2005. A\&A, 444, 347

\item{}
Velikhov, E.P. 1959. Sov. Phys. JETP,  9, 995

\end{thebibliography}
\end{document}